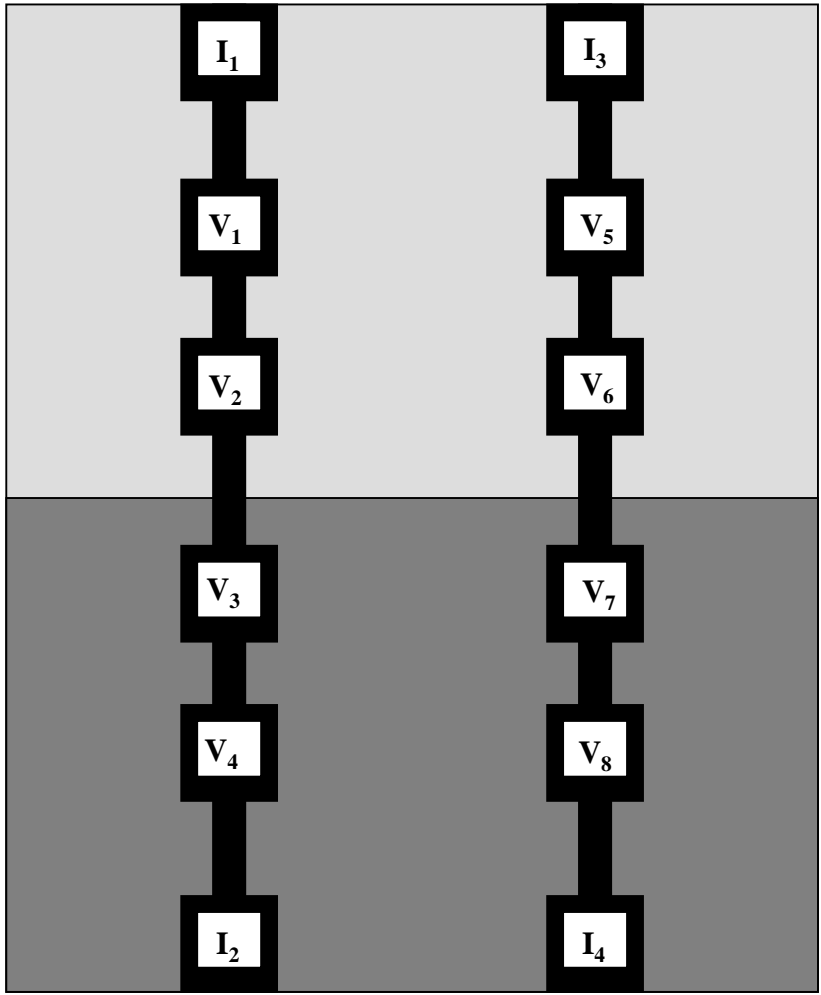

**Fig.1**

S.P.Pai *et.al*

**Title**: Self Injection length in $La_{0.7}Ca_{0.3}MnO_3$-$YBa_2Cu_3O_{7-\delta}$ ferromagnet-superconductor multi layer thin films

**Authors**: S.P. Pai[1], S. Wanchoo[2], S.C. Purandare[1], T. Banerjee[1],

P.R. Apte[1], A.M. Narsale[2] and R. Pinto[1]

[1]Tata Institute of Fundamental Research, Mumbai, India
[2]Dept. of Physics, Mumbai University, Mumbai

**Comments:** Six (6) pages, one figure in .ps format

**Subject class:** Superconductivity

We have carried out extensive studies on the self-injection problem in barrierless heterojunctions between $La_{0.7}Ca_{0.3}MnO_3$ (LCMO) and $YBa_2Cu_3O_{7-\delta}$ (YBCO). The heterojunctions were grown in situ by sequentially growing LCMO and YBCO films on <100> $LaAlO_3$ (LAO) substrate using a pulsed laser deposition (PLD) system. YBCO micro-bridges with 64 μm width were patterned both on the LAO (control) and LCMO side of the substrate. Critical current, $I_c$, was measured at 77K on both the control side as well as the LCMO side for different YBCO film thickness. It was observed that while the control side showed a $J_c$ of ~2 x $10^6$ A/ $cm^2$ the LCMO side showed about half the value for the same thickness (1800 Å). The difference in $J_c$ indicates that a certain thickness of YBCO has become 'effectively' normal due to self-injection. From the measurement of $J_c$ at two different thickness' (1800 Å and 1500 Å) of YBCO both on the LAO as well as the LCMO side, the value of self-injection length (at 77K) was estimated to be ~900Å. To the authors' best knowledge, this is the first time that self-injection length has been quantified. A control experiment carried out with $LaNiO_3$ deposited by PLD on YBCO did not show any evidence of self-injection.

## 1. Introduction

A study of spin-polarized electron transport from ferromagnetic materials to superconductors is important for the understanding of spin dependent electronic properties and for the realization of possible spin injection devices. Although, pioneering experiments were carried out by Tedrow and Meservey [1], and a theoretical understanding of the injection of spin polarized carriers into superconductors was first described by Aronov [2], it is only recently that the importance of magnetoelectronics has been realized with the discovery of the colossal magnetoresistive (CMR) materials [3,4]. The electrical transport in these CMR manganites such as $La_{0.7}Ca_{0.3}MnO_3$ involves spin-polarized carriers whose degree of polarization is close to unity. Therefore, as these materials have good structural compatibility with high $T_c$ superconductors (HTSC), and are a rich reservoir of spin-polarized charges, they can be used for spin injection studies. Injection of spin-polarized electrons into a superconductor with energy greater than the energy gap induces non-equilibrium state in the superconductor by creating non-equilibrium population of spins in the material, both by scattering and pair breaking phenomenon. The polarized conduction electron's spin orientation is eventually randomized by spin scattering events. The characteristic length at which the spins of the polarized carriers is randomized by scattering events in the superconductor at CMR-superconductor interface is known as the spin diffusion length.

Here we report our observations on the effects of self-injection of spin-polarized carriers on specially fabricated devices which are essentially a heterostructure consisting of a layer of $YBa_2Cu_3O_{7-\delta}$ (YBCO), grown over $La_{0.7}Ca_{0.3}MnO_3$ (LCMO). The critical current of the YBCO film grown on LCMO layer is found to be much lower than the critical current of the YBCO film grown on bare LAO, thereby indicating the effect of self injection of spin polarized electrons from LCMO into YBCO. By achieving a good control in the quality of the constituent layers, we report the successful experimental measurement of the self-injection length in perovskite LCMO-YBCO heterostructures for the first time after carrying out extensive studies on the self-injection problem in barrierless heterojunctions. When the LCMO layer is replaced by a $LaNiO_3$ (LNO) layer, which behaves as a normal metal, it is observed that its effect on the critical current of the YBCO film is negligible since there are no polarized spins so as to cause self-injection.

## 2. Experimental details

In order to determine the self-injection length of spin-polarized carriers from a CMR ferromagnet (LCMO) into a superconductor (YBCO), micro-bridge devices were fabricated on LCMO/YBCO heterostructures with YBCO thickness of 1500Å and 1800Å. The thickness of the LCMO layer was ~500 Å in both the devices. In order to make sure that the YBCO film on LCMO is of the same high quality both films were deposited one after other *in situ*. X-ray diffraction studies showed that both YBCO and LCMO/YBCO multi-layers were highly c-axis oriented. In practice, we used a device geometry shown in Fig. 1. First, a 500 Å thick LCMO layer was grown at $800^\circ$C by Pulsed laser deposition (PLD) on one half of a 10 mm×10 mm×0.5 mm LAO substrate. The other half was masked using another clean LAO substrate. The LAO mask was then dropped *in situ* and YBCO film was then deposited on the entire substrate at $800^\circ$C. This was done to obtain a high quality (barrierless) interface between LCMO/YBCO. At the same time it also gives a control YBCO film on bare LAO. Two sets of micro-bridges, each 64μm wide, were then patterned using UV photolithography and chemical etching as shown in Fig. 1. To obtain different YBCO film thicknesses on the same substrate/in the same run, the YBCO film was grown slightly off-axis. This gave micro-bridges of 1800 Å and 1500 Å thickness. Au contacts were then deposited by PLD (using reverse lift-off technique). Referring to Fig. 1, two contacts each ($I_1$-$I_2$ and $I_3$-$I_4$) were used for current and two each ($V_1$-$V_2$, $V_5$-$V_6$ and $V_3$-$V_4$, $V_7$-$V_8$) for voltage measurements (for micro-bridges on control YBCO and YBCO/LCMO). Very thin copper wires were attached to each of the gold contacts on the substrate using indium solder. The LAO substrate with the micro-bridges was then mounted on a $LN_2$ insert and the measurements were made at 77 K by dipping the insert into $LN_2$ dewar.

## 3. Results and Discussion

The superconducting transition temperature, $T_c$, and critical current, $I_c$, of YBCO micro-bridges were measured on each arm of the device for two YBCO film thicknesses. The $T_c$ of the YBCO films was 90K on both the control as well as LCMO side. However, the critical current density, $J_c$, of YBCO micro-bridges on the LAO side was higher than that on the LCMO side. The actual values of measured $J_c$ were $2 \times 10^6$ A/cm$^2$ on control side and $1 \times 10^6$ A/cm$^2$ on the LCMO side (for the 1800 Å thick bridge). Assuming that properties are uniform along thickness axis, this indicated that half the thickness of YBCO had effectively become normal (due to self-injection). The thickness being 1800 Å, the self-injection length appears to be ~900 Å. To further check on this, $J_c$ was measured on the 1500 Å thick bridges as well. Taking out 900 Å from 1500 Å leaves effectively 600 Å of superconducting YBCO (on LCMO side). The measured value of $J_c$ agreed with this value ($0.7 \times 10^6$ A/cm$^2$). This result was also confirmed by depositing YBCO films of thickness $\leq 800$ Å on LCMO film. These films did not show any $T_c$.

On the other hand, when a control experiment was performed by depositing *in situ,* a 500 Å thick normal metal (LNO) directly on YBCO film on LAO substrate, the YBCO film showed a $T_c$ of 90 K, even when its thickness was < 500 Å. The above experiments show that a layer of YBCO, ~900 Å in thickness, directly in contact with the LCMO ferromagnetic layer, has become non-superconducting. The possibility of an insulating barrier layer between LCMO and YBCO is ruled out as both the layers were grown *in situ* with the YBCO having a good lattice match to LCMO. In the worst case, only a monolayer or two of YBCO directly in contact with the LCMO layer may be weakly superconducting probably due to oxygen deficiency at this interface. Further, the fact that a normal metal, LNO, grown *in situ* on YBCO does not affect the superconductivity of YBCO even when its thickness is <500 Å (well below the estimated spin injection thickness) indicates that the loss of superconductivity cannot be linked to a thick insulating barrier at the interface. Hence it is clear that the YBCO layer, ~900 Å in thickness, is rendered non-superconducting due to the self injection of spin polarized carriers at the LCMO/YBCO interface.

We shall now discuss the implication of the suppression in $J_c$ in the case of a LCMO-YBCO heterostructure. Self-injection of carriers (as opposed to external injection due to applied electric field) occurs when two materials are brought in contact with each other. In the present experiment, YBCO, a superconductor with significant number of quasi-particles at 77 K, is in contact with ferromagnetic LCMO layer with a high density (near unity) of spin-polarized carriers. An important feature of the high $T_c$ superconductors is that their order parameters have unconventional symmetry [5] as opposed to the ordinary metallic superconductors exhibiting s-wave symmetry. The order parameter and pairing state of these high $T_c$ materials is anisotropic, exhibiting d-wave symmetry [6,7] Hence, the gap in the *k*-space distribution function of the electrons goes to zero or is small in the vicinity of the four nodes. These nodal or quasi-nodal directions are easy directions for quasi-particle excitation (low energy excitations) and one expects a significant population of quasi-particle excitations even at very low temperature. Moreover, finite impedance exists at the interface between the LCMO and YBCO, which builds up a chemical potential [8], when a current flows across the interface. Self-injection of thermally induced quasi-particles occurs from YBCO to LCMO, which in turn sustains and assists further diffusion of quasi-particles. The ferromagnetic LCMO, having near perfect spin-polarization, allows only those quasi-particles to diffuse across the interface whose spins are parallel to those of the majority carriers in the LCMO, leaving behind spin-polarized quasi-particles in the superconductor. This induces further pair breaking. Interestingly, suppression of $J_c$ is found to be a function of the thickness of the intermediate insulating layer [8]. In our case, a weakly superconducting layer that might be just a few monolayers thick is probably formed at the interface of the LCMO-YBCO. A suppression of $J_c$ is observed in this case and is explained by the self-injection of quasi-particles from YBCO to the LCMO layer [8,9].

## 4. Conclusions

In conclusion, we have fabricated and characterized heterostructures of a ferromagnet (LCMO) and a high $T_c$ superconductor (YBCO) by PLD. The critical current of the YBCO film over the LCMO layer is found to be suppressed due to self-injection of spin-polarized quasi-particles across the interface. The resultant increase in the number density of spin-polarized quasi-particles in YBCO enhances the pair breaking effect, thereby suppressing the critical current in YBCO. After carrying out rigorous studies on barrierless junctions and achieving good control on the quality of the deposited layers, we were successful in estimating the self-injection length in these heterostructures. The spin injection length for this bilayer structure is found to be 900 Å. A control experiment carried out by replacing the LCMO layer with LNO shows no suppression of $J_c$, thus corroborating the fact that the self-injection of spin-polarized carriers is solely responsible for a decrease of $J_c$ at low temperatures in a LCMO-YBCO heterostructures.

**Figure caption:**

Fig.1: Device structure and electrical contact configurations for measurements of $I_C$. Contacts $I_1$- $I_2$, $I_3$-$I_4$ were used as current contacts and $V_1$ to $V_8$ were used as voltage contacts for measuring the Ic's of micro-bridges on LAO side as well as LCMO side.